\documentclass[iop]{emulateapj}
\usepackage[T1]{fontenc}
\usepackage{ae,aecompl}

\usepackage{graphicx}	
\usepackage{amsmath}	
\usepackage{amssymb}	

\shorttitle{Disintegration of Berkeley 17}
\shortauthors{ }

\begin{document}

\title{
	Disintegration of the Aged Open Cluster Berkeley~17
	}

\author{
	Souradeep Bhattacharya\altaffilmark{1}, Ishan Mishra\altaffilmark{2}, Kaushar Vaidya\altaffilmark{1}, 
	W.~P. Chen\altaffilmark{3}
       }
\email{f2012553@pilani.bits-pilani.ac.in}

\altaffiltext{1}{Department of Physics, Birla Institute of Technology and Science, Pilani 333031, Rajasthan, India}
\altaffiltext{2}{Indian Institute of Technology Guwahati, Guwahati 781039, Assam, India}
\altaffiltext{3}{Graduate Institute of Astronomy, National Central University, 300 Jhongda Road, Jhongli 32001, Taiwan}

\begin{abstract}
We present the analysis of the morphological shape of Berkeley\,17, the oldest known open cluster ($\sim10$~Gyr), using a probabilistic star counting of Pan-STARRS point sources, and confirm its core-tail shape, plus an antitail, previously detected with the 2MASS data. The stellar population, as diagnosed by the color-magnitude diagram and theoretical isochrones, shows many massive members in the cluster core, whereas there is a paucity of such members in both tails. This manifests mass segregation in this aged star cluster with the low-mass members being stripped away from the system. It has been claimed that Berkeley\,17 is associated with an excessive number of blue straggler candidates.  Comparison of nearby reference fields indicates that about half of these may be field contamination.  
\end{abstract}


\section{Introduction} 

Most, and likely all, stars are formed in a clustered environment out of molecular clouds \citep{lad03}. Those surviving cloud dispersal and remaining gravitationally bound are seen as open clusters, with tens to thousands of member stars. A cluster of equal-mass stars is dynamically relaxed on a time scale of $\tau_{\rm relax}\approx \tau_{\rm cross} \,0.1 N/ \ln N$ \citep{bin87}, where $\tau_{\rm cross} \approx D/v$ is the crossing time for a system with a characteristic size $D$ and typical velocity $v$, and $N$ is the total number of stars.  {As a result of energy equipartition among stars of differing mass, more massive stars take on a smaller velocity dispersion and ``sink" to the center of the cluster, whereas lower-mass stars occupy a larger volume as their higher velocities carry them farther from the cluster center.} One of the consequences of this ``mass segregation" process is the lowest-mass members become most vulnerable to be ejected out of the system \citep[e.g., see][]{mat84}. This ``stellar evaporation", with an $e$-folding time scale $\tau_{\rm evap} \approx 100 ~\tau_{\rm relax}$ \citep{shu82,bin87}, leads to a continuing decrease of the total mass, and hence the gravitational binding energy, of the cluster.  Any external disturbance, such as the tidal force from nearby giant molecular clouds or star clusters, passages through Galactic spiral arms or disks, or shear force by Galactic differential rotation, exacerbates the disintegration of the cluster.  A recently dissolved system in the solar neighborhood could be recognized as a ``moving group" as the then-members still share systemic kinematics and distances \citep{zuc04}. Eventually, the escaped stars supply the disk field population.  

Berkeley\,17 (RA=05:20:37, DEC=+30:35:12, J2000) was first identified by \citet{sw62}. Located near the Galactic anti-center ($\ell=175\degr.657$, $b=-3\degr.649$), with a metallicity ${\rm[Fe/H]}\approx-0.33$ \citep{fri02}, the cluster with an age $\sim10$~Gyr is considered among the oldest Galactic open clusters \citep{kru06,sal04,phe97,kal94}, rivaling globular clusters. While \citet{bra06} found a slightly lower age of 8.5--9~Gyr, these authors did not rule out an older age up to $\sim12$~Gyr. For the distance, \citet{kha13} found a distance of 1800~pc from the sun, and \citet{kal94} placed the cluster beyond the disk, at 4400~pc.  In this work, we adopt the distance of 2700~pc from the sun with a reddening $E(B-V) = 0.7$ as found by \citet{bra06} and \citet{phe97}. The cluster has a mean proper motion $\mu_{\rm \alpha} \cos\delta = 3.60$~mas~yr$^{-1}$, $\mu_{\rm \delta} =-3.62$~mas~yr$^{-1}$ \citep[UCAC4;][]{dia14}.

Our study aims to investigate how an aged disk cluster like Berkeley\,17 survives the disruption processes. \citet{che04} applied probabilistic star counting of sources from the Two Micron All Sky Survey (2MASS), an all-sky survey in near-infrared wavelengths \citep{skr06}, and inferred an elongated shape for this cluster, containing some 370 members within a mean radius of $8\farcm19$, with a noticeable tail pointing to the Galactic disk. The angular extent of the tail is comparable to the core of the cluster itself, $\sim6\arcmin$--$7\arcmin$, or about 5~pc in projected length given the distance to Berkeley\,17. Interestingly, the structural analysis \citep{che04} also suggested an antitail, suggestive of tidal origin, as opposed to stellar debris trailed in the cluster's orbital motion.

\begin{figure}[htb]
	\centering
	\includegraphics[height=0.85\columnwidth,angle=0]{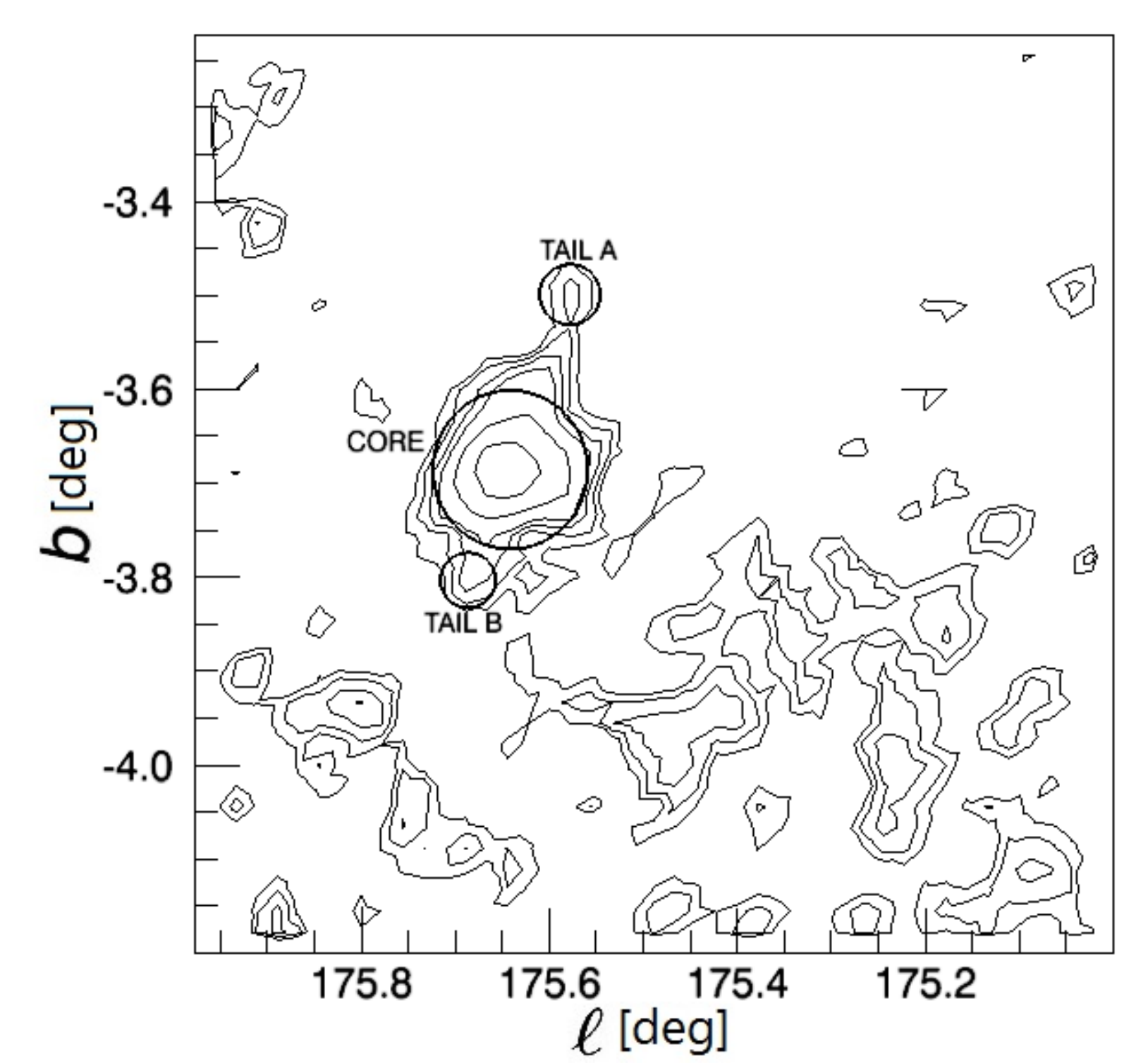}
	\caption{The contour plot shows the core-tail morphology of Berkeley 17.  The contour levels indicate the clustering parameters varying from $0.58$ to $0.75$. The levels are not in equal spacing so as to bring up the optimal clarity. The analyzed regions, the core and two tidal tails, have been marked.}
	\label{fig:contour}
\end{figure}

In this work, we present the analysis of the stellar population in the tails, in comparison to that in the core, in order to diagnose stripping of low-mass member stars from the cluster. Given the distance to the cluster, the proper motion data cannot be used readily to discriminate members against field stars. Thus, we utilized stellar photometric data obtained from the Panoramic Survey Telescope And Rapid Response System \citep[Pan-STARRS;][]{hod04,ton12} for our analysis. Berkeley\,17 in many ways resembles a globular cluster.  For example, in addition to a relatively poor metallicity, it is known to have a rich population of blue stragglers \citep{ahu07}, which are stars lying above the main-sequence turn-off in a star cluster's color-magnitude diagram (CMD). In this work, we diagnose how the previously identified blue stragglers could have been confused with field stars.

In Section~\ref{data}, we describe the Pan-STARRS data used in our work. Section~\ref{mor} relates to our analysis of the data to understand the morphology of Berkeley\,17 and its mass distribution using field decontaminated CMDs. We also present our analysis of blue straggler candidates in this section. We discuss our results in Section~\ref{dis}.

\section{Data Description}\label{data}

The prototype Pan-STARRS (PS1) used a 1.8 meter telescope, located atop Haleakala, Maui, USA \citep{kai10}, and its 1.4 gigapixel camera \citep{ton08} to image the sky through a set of five broadband filters, termed $grizy_{\rm P1}$ \citep{ton12}. The PS1 filters differ from those of the SDSS \citep{aba09} in that the $g_{\rm P1}$ filter extends 20~nm redward of $g_{\rm SDSS}$ for greater sensitivity and lower systematics for photometric redshift estimates, and there is no corresponding $y$ filter in SDSS.  For the details of the PS1 surveys and latest data products, see \citet{cha16}. PS1 provides deeper photometric data than 2MASS but has a saturation limit around $\sim14$~mag \citep{mag13}. 

For this work, only objects with a clear detection in every PS1 band were included in the analysis.  While PS1 had observations at multiple epochs, for the results reported here only the average magnitude in each band was used. The PS1 sample consists of stars within a field of $1\degr \times 1\degr$ around Berkeley\,17, and with photometric uncertainties less than 0.1~mag in both $g_{\rm P1}$ and $y_{\rm P1}$.

\begin{figure}[htb]
	\centering
	\includegraphics[height=0.85\columnwidth,angle=0]{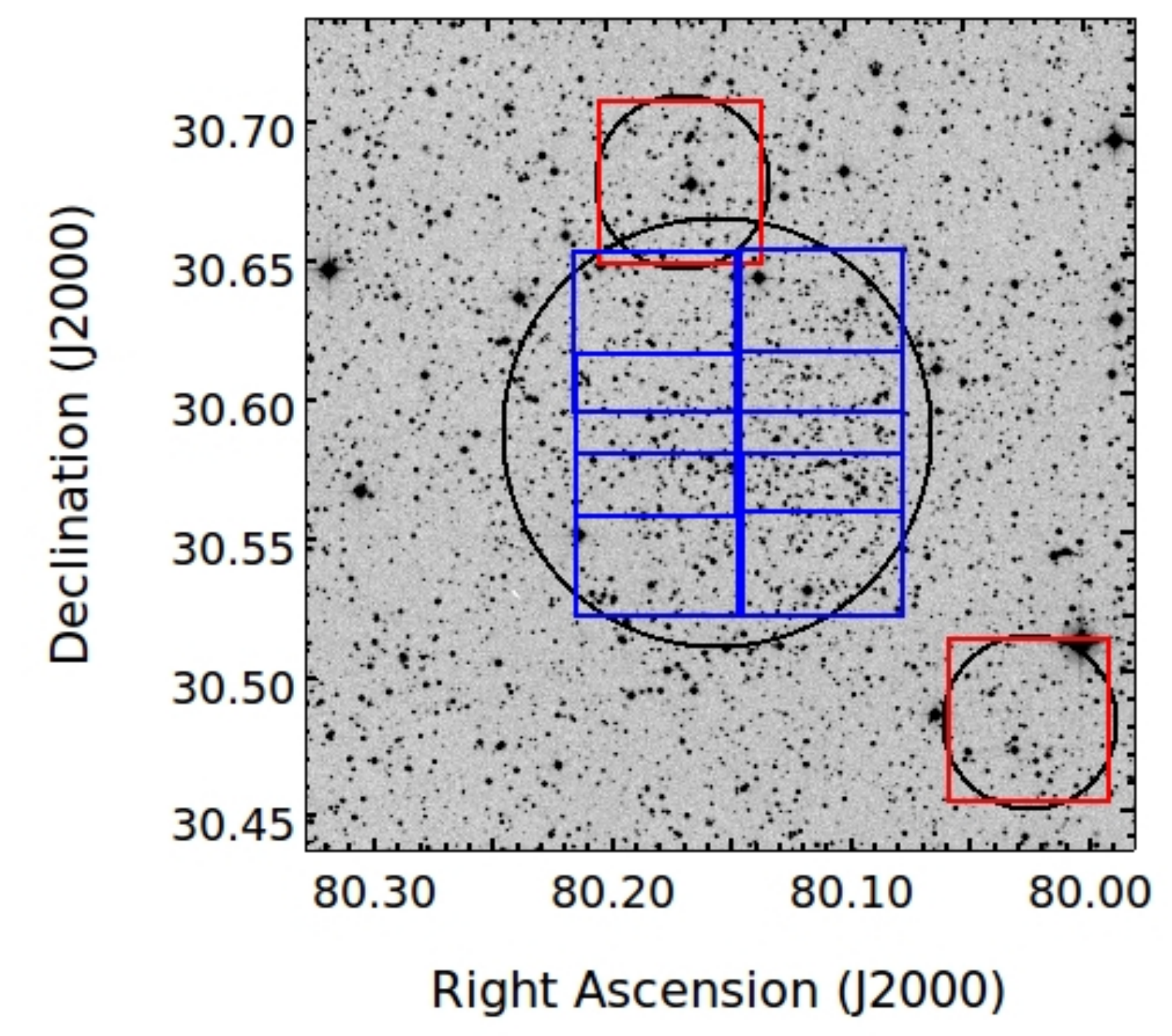}
	\caption{An $18\arcmin \times 18\arcmin$ DSS red image showing the core region as defined in Section~\ref{cm} marked as a big circle and the two tail regions marked as small circles. The six squares of size $3\farcm5$ enclosed within the large circle show the extent of the six $g_{\rm P1}$ images used for completeness determination of the core whereas the two squares of $3\farcm5$ size enclosing the two small circles show the area of the $g_{\rm P1}$ images used for completeness determination of the two tails.}
	\label{fig:regions}
\end{figure}

\section{Data Analysis} \label{mor}

\subsection{Cluster Morphology} \label{cm}

For every PS1 source within our region of analysis, the clustering parameter was computed, prescribed in \citet{che04}, as $P(i)= (N_{\rm t}(i) - N_{\rm f}(i)) /N_{\rm t}(i) $, where $N_{\rm t}(i)$ is the total number of neighbouring stars for star $i$ within a specified ``neighborhood radius", and $N_{\rm f}$ is the average number of field stars within the same radius in a field displaced from the cluster yet representative of the field star population.  Effectively, $P(i)$ is a measure of the spatial probability of a cluster member, being close to unity when the cluster density is high relative to the background density, and nearly null in a background field.  The surface number density of stars is then calculated by summation of the clustering parameter of every star in a sky grid. For Berkeley\,17, the neighborhood radius was chosen to be $0\fdg03$, yielding an average $N_{\rm f}=50$ with $N_{\rm t}$ reaching up to 420 in the densest part of the cluster.

It is reassuring that in the analysis using PS1 data, the tail-like structure detected with 2MASS \citep{che04} was recovered, as seen in Figure~\ref{fig:contour}. The approximate centers of the tails are located, respectively at $(\ell, b)=(175\fdg592, -3\fdg591)$ (hereafter tail~A), and $(\ell, b)=(175\fdg682, -3\fdg804)$ (hereafter tail~B).
We define a radius of $4\farcm648$ for the ``core" of the cluster, and a radius of $1\farcm897$ for each of the two tails A and B. Our goal was to compare the stellar properties in the cluster core and in the tails. 

\begin{figure*}[htb]
	\centering
	\includegraphics[height=0.85\columnwidth,angle=0]{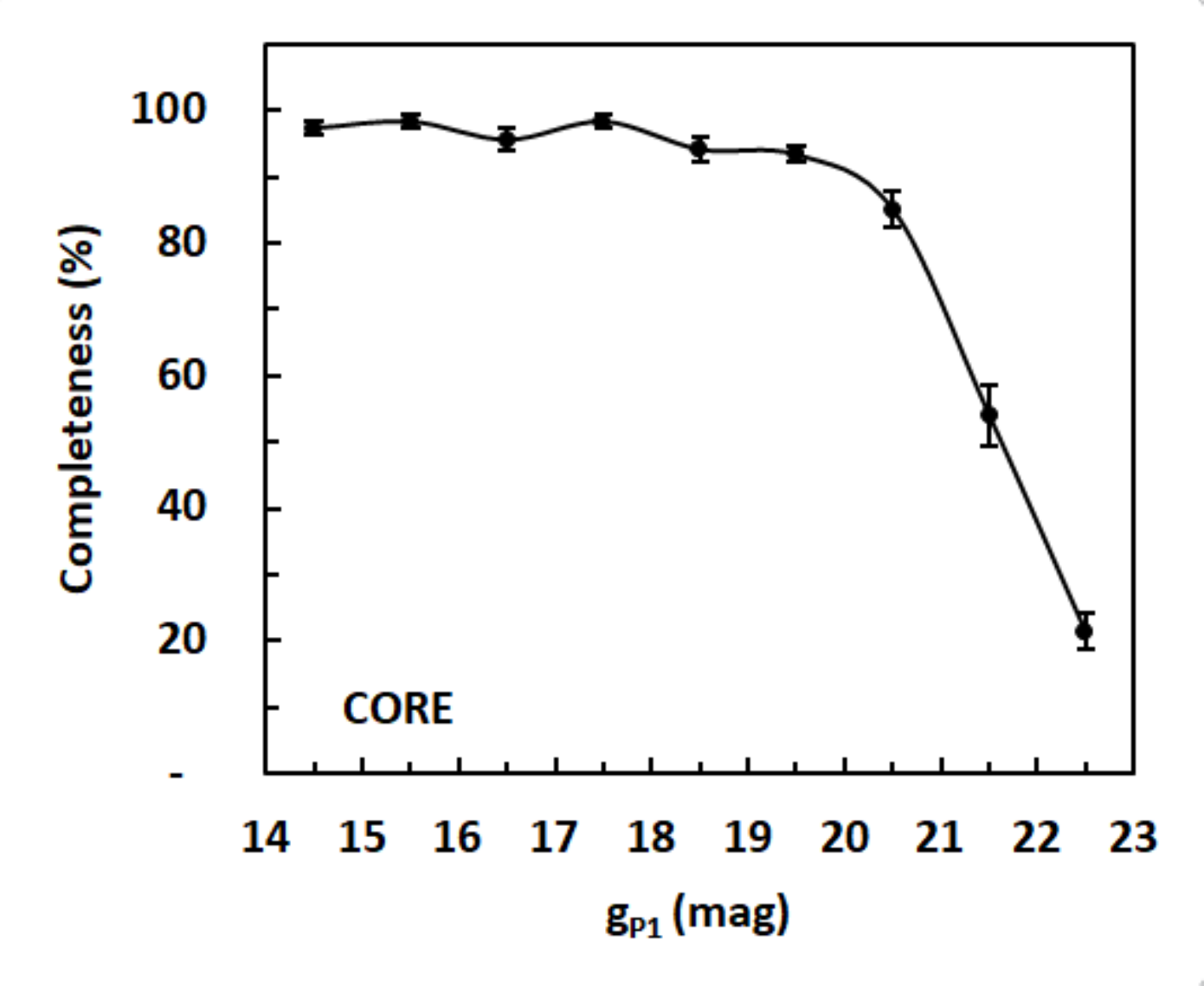}
	\includegraphics[height=0.85\columnwidth,angle=0]{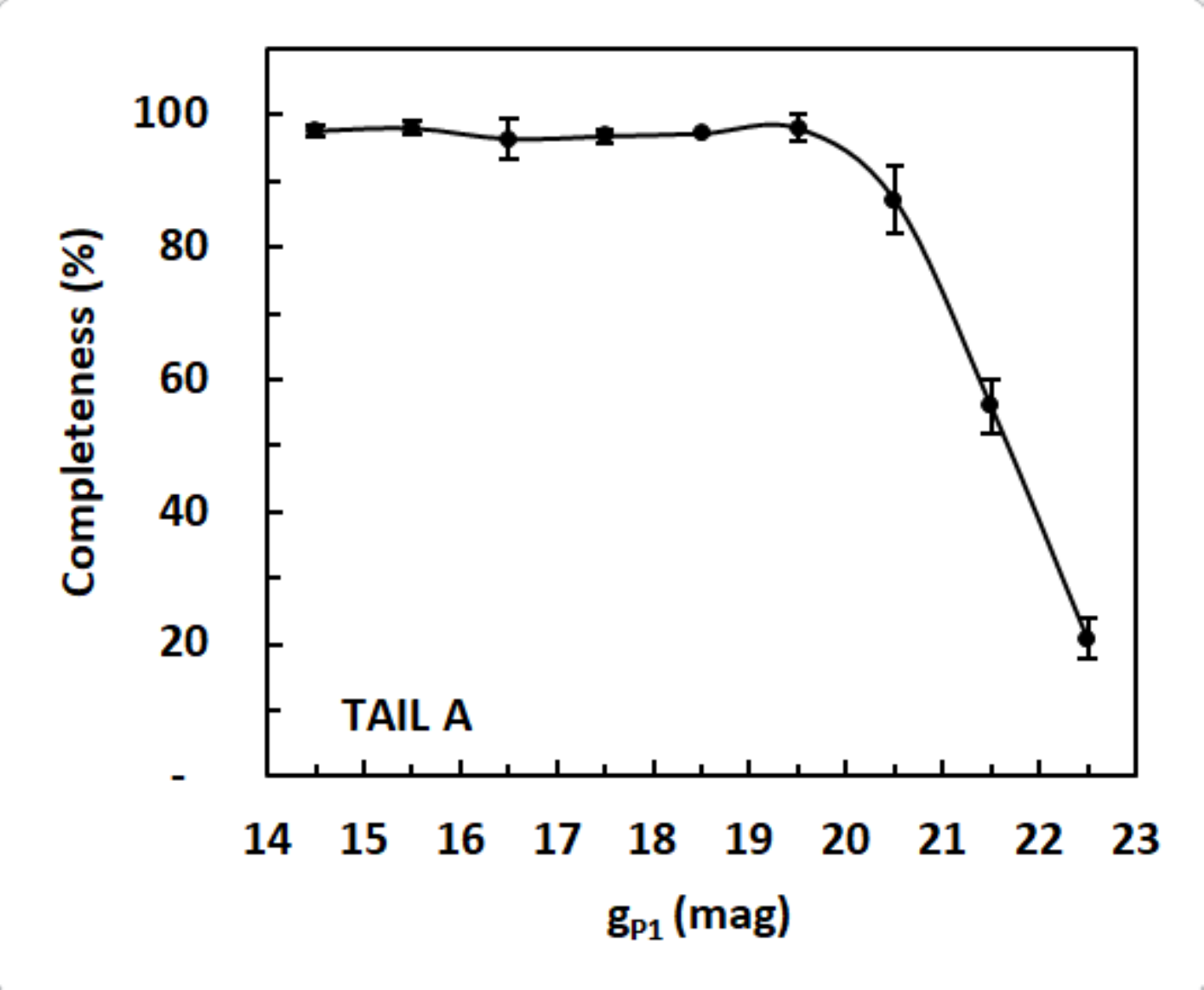}
	\includegraphics[height=0.85\columnwidth,angle=0]{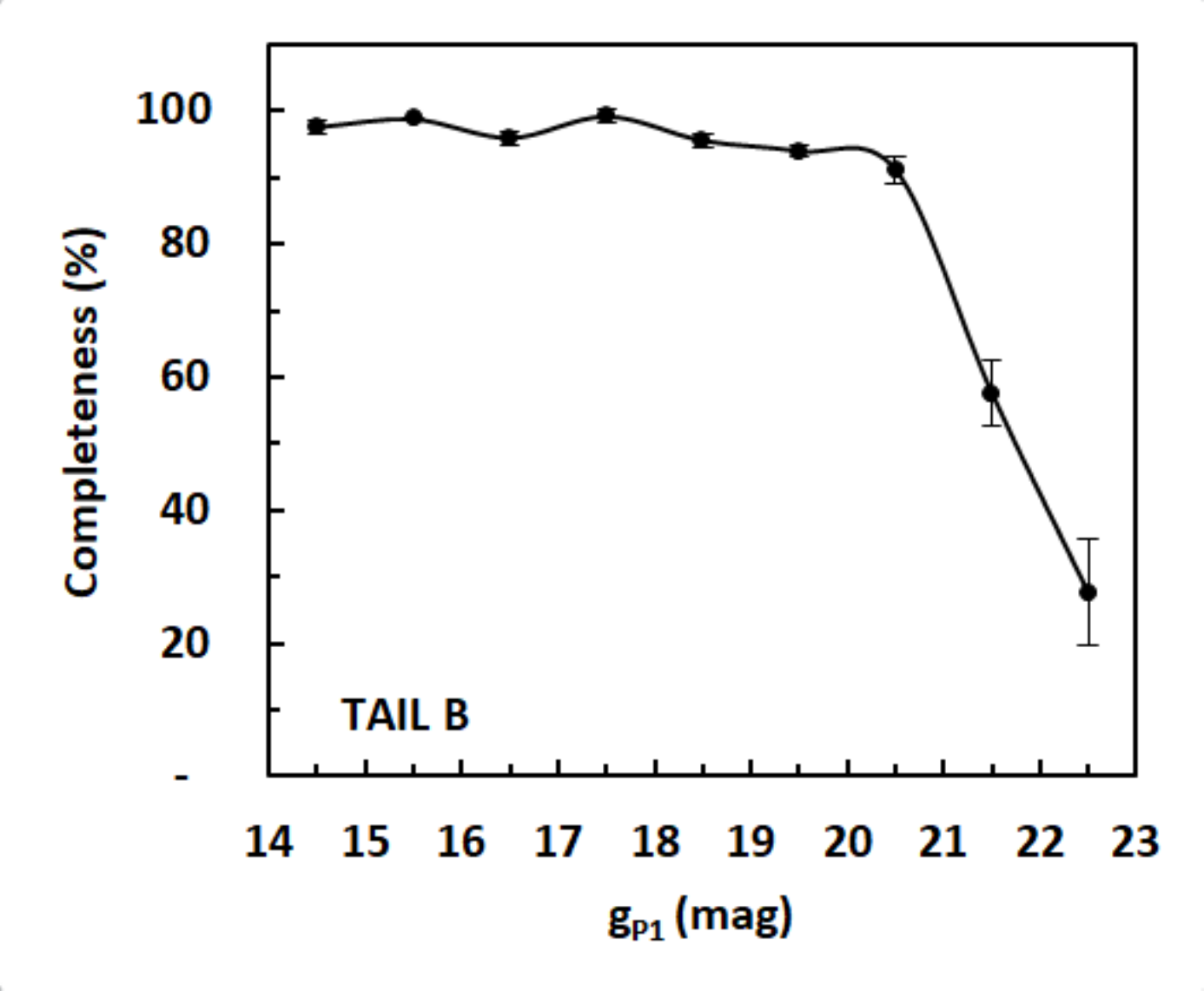}
	\includegraphics[height=0.85\columnwidth,angle=0]{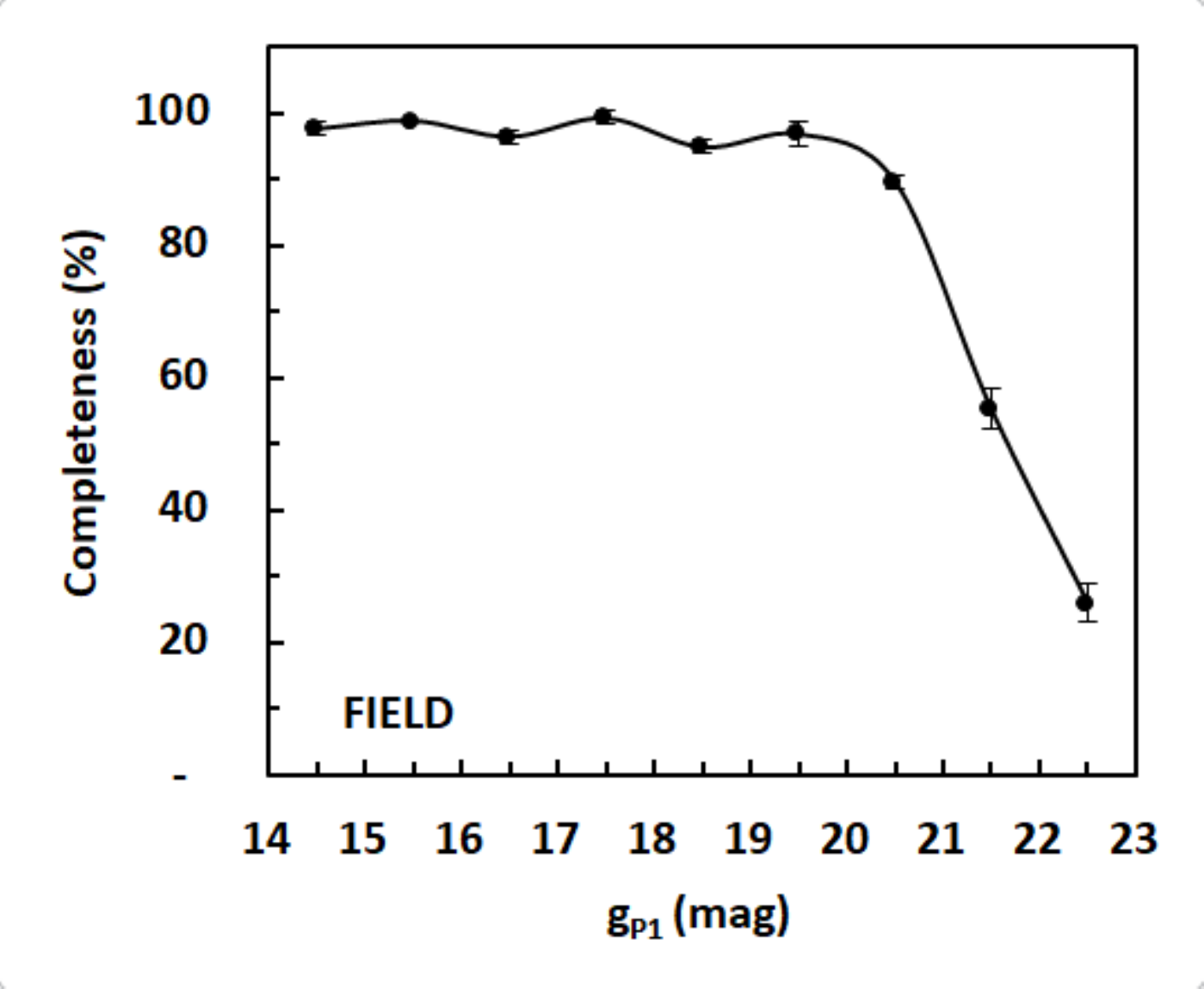}
	\caption{The completeness curves for the cluster core, the two tails, and the field used for decontamination.}
	\label{fig:comp}
\end{figure*}

\subsection{Photometric Completeness}\label{comp}

To determine the photometric completeness limits of the data for our region, and to particularly correct for any significant differences in the photometric depth in the core of the cluster with respect to the two tails, we made use of the ADDSTAR task of the Image Reduction and Analysis Facility (IRAF)\footnote{http://iraf.noao.edu/}.  {We downloaded $g_{\rm P1}$ band images for the core and the two tails using the PS1 image cutout facility\footnote{http://ps1images.stsci.edu/cgi-bin/ps1cutouts}. The complete core image was not available in a single cutout. We downloaded eight images, each of box size $3\farcm5$, six for the core and one image each for the tails. Figure~\ref{fig:regions} shows the regions of these images as well as the entire core and the two tails as defined in Section~\ref{cm}, marked on an $18\arcmin \times 18\arcmin$ DSS red image of Berkeley 17.}

 {We first performed PSF photometry on all eight images. After that, we used ADDSTAR task to add 50 stars at random locations in the images for a bin size of one magnitude, for the entire range of 14--23 mag. We generated five lists of randomly placed stars for each magnitude bin, and added those stars to create five new sets of images for each of the eight images. We then performed photometry on each ADDSTAR generated image, with the same procedure to process the original observed images, to determine the retrieval rate for fake stars. The mean of the five individual retrieval rates for a given magnitude bin was computed to determine the completeness in the given magnitude bin. We thus determined completeness curves for each of our eight images. For the core, we found no significant variation of completeness within the six subregions. We adopted the mean value of the completeness for each magnitude bin of the six subregions of the core as the completeness of the core.} The resultant completeness curves in the $g_{\rm P1}$ band for the core and the two tails are shown in Figure~\ref{fig:comp}. The completeness for the core, tail A and tail B, is $85\%$, $87\%$ and $91\%$ respectively, in the 20--21 mag bin and is $54\%$, $56\%$ and $58\%$ respectively, in the 21--22 mag bin. The core is marginally less complete photometrically because of a higher stellar density, but the completeness limits in the core and in the two tails are comparable.  {Figure~\ref{fig:comp} also shows the completeness curve for the field used for decontamination (details in Section~\ref{cmd}), which is also found to be comparable to that of the cluster regions.}


\begin{figure}[htb]
	\centering
	\includegraphics[height=0.75\columnwidth,angle=0]{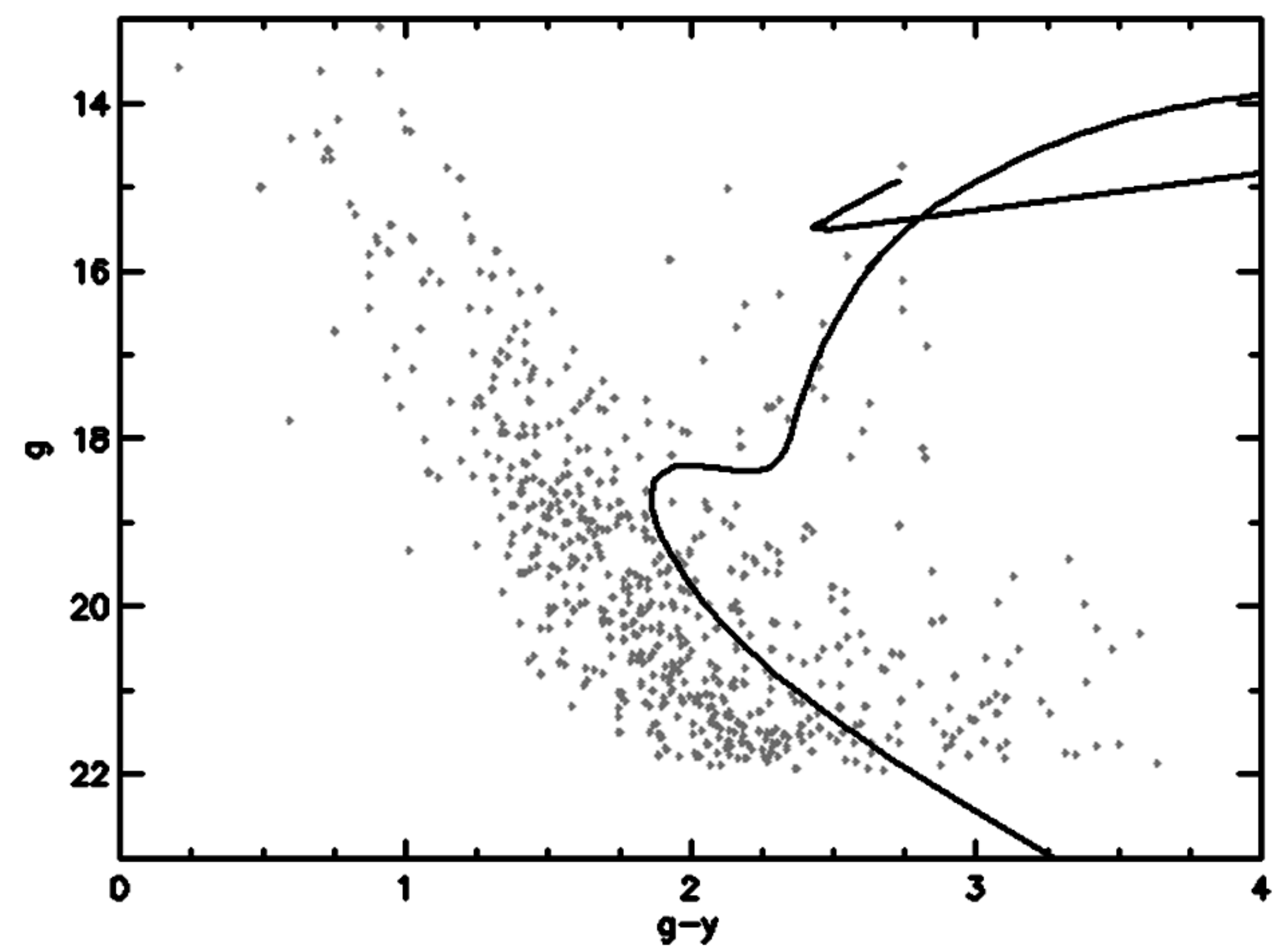}
	\includegraphics[height=0.75\columnwidth,angle=0]{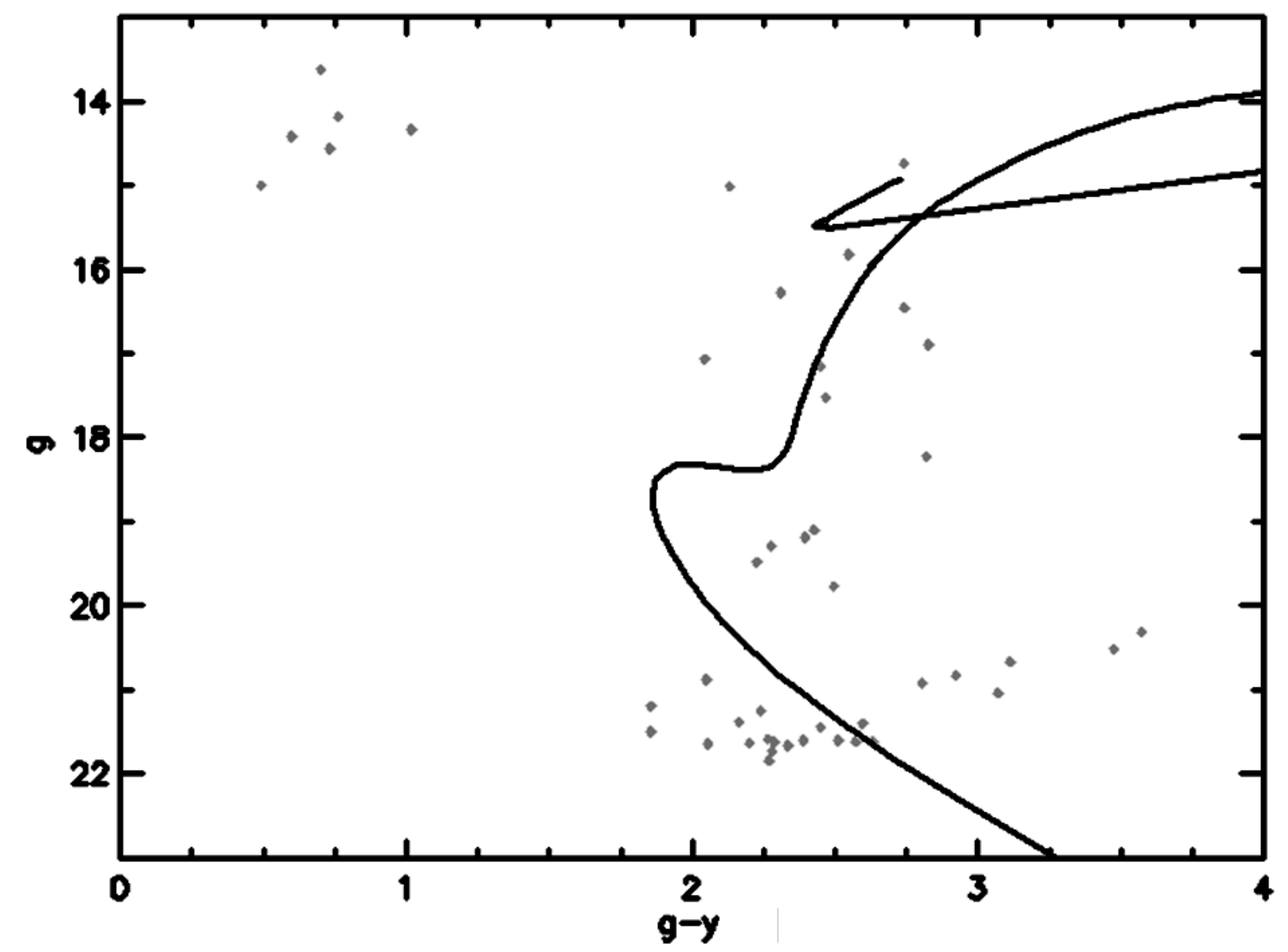}
	\caption{(Top) the field CMD and (bottom) the field CMD cleaned by the check field. The solid line corresponds to the 10-Gyr isochrone appropriate for Be\,17.}
	\label{fig:fieldclean}
\end{figure}

\begin{figure}[htb]
	\centering
	\includegraphics[height=0.75\columnwidth,angle=0]{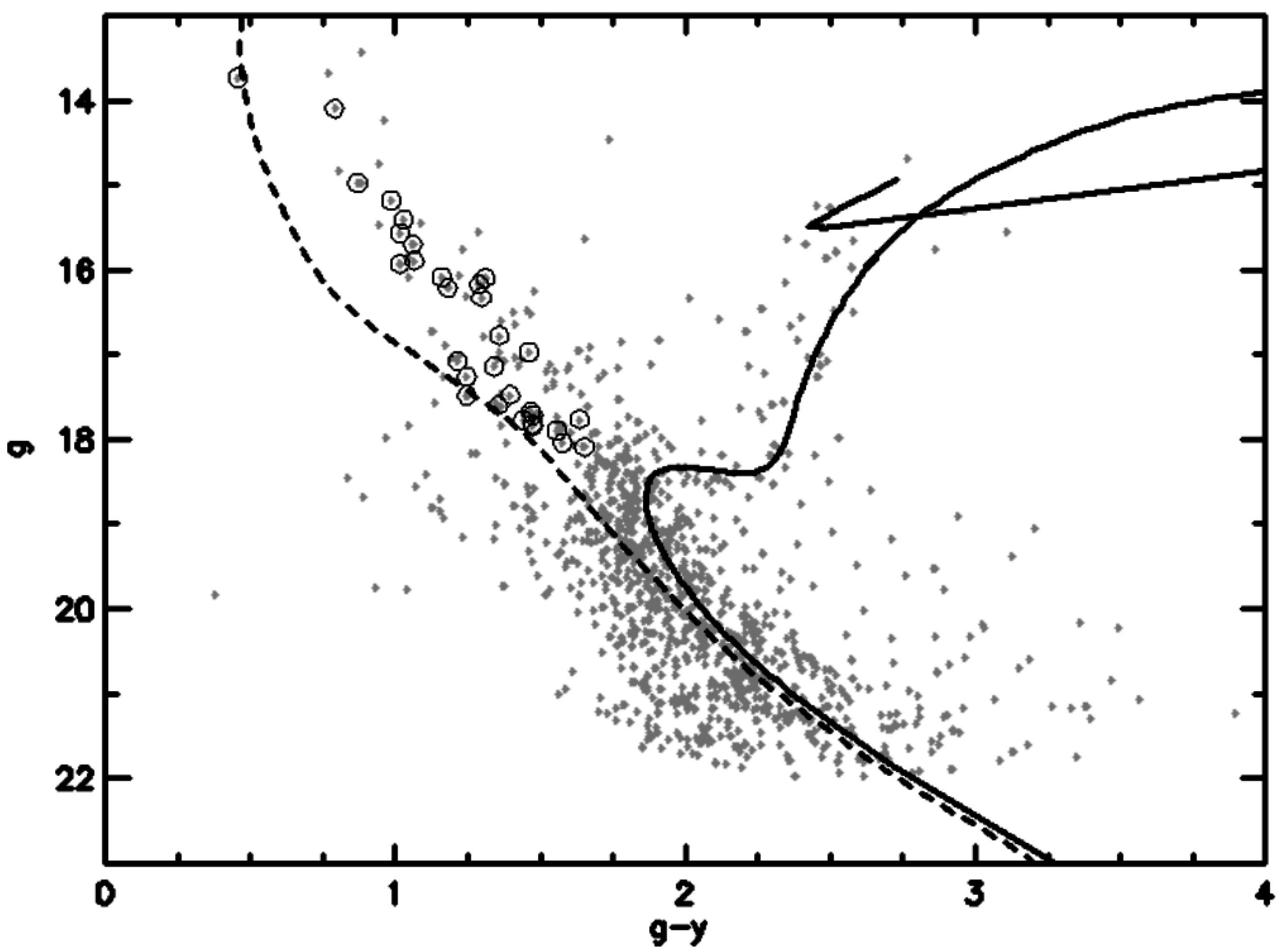}
	\includegraphics[height=0.75\columnwidth,angle=0]{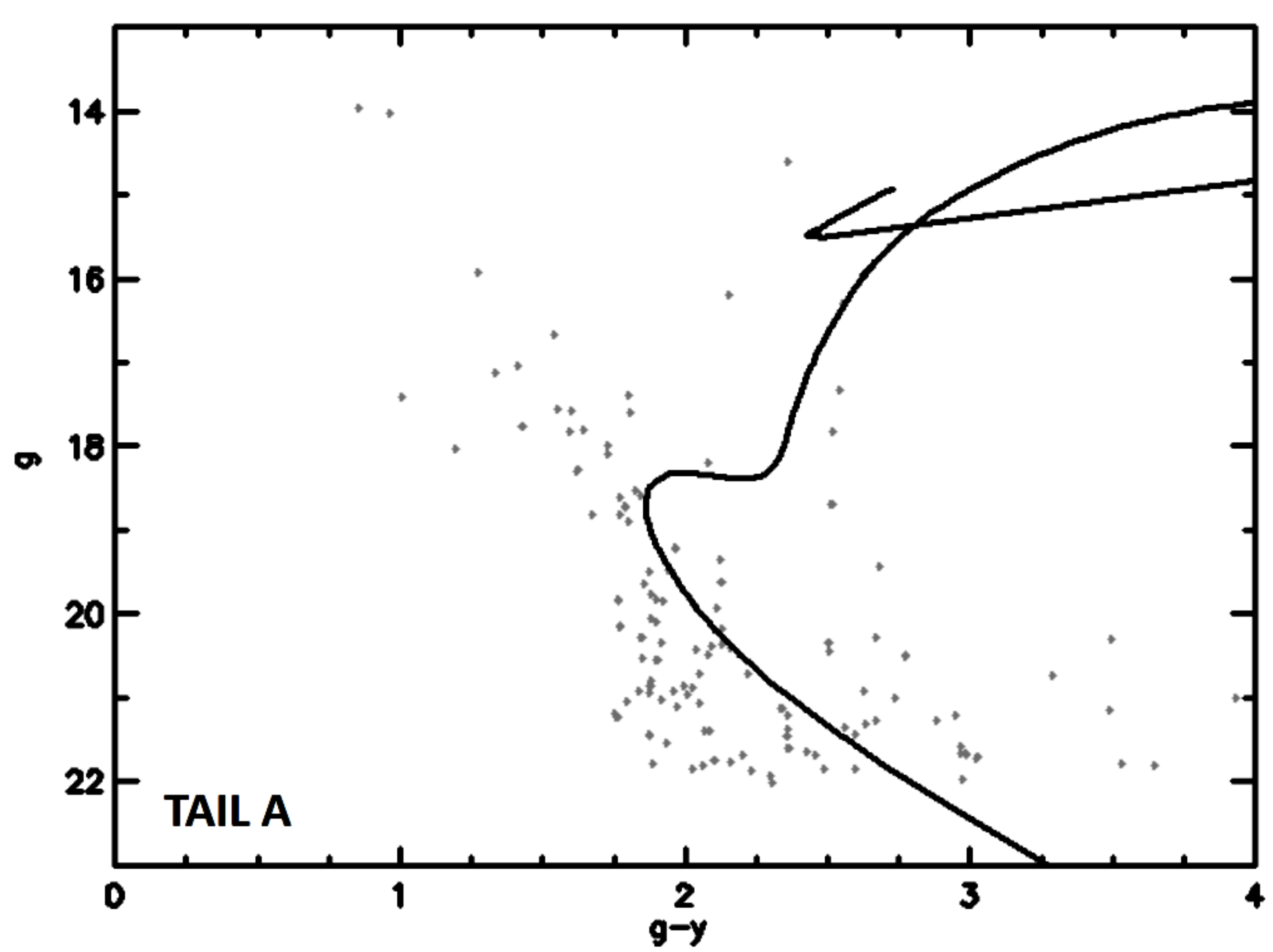}
	\includegraphics[height=0.75\columnwidth,angle=0]{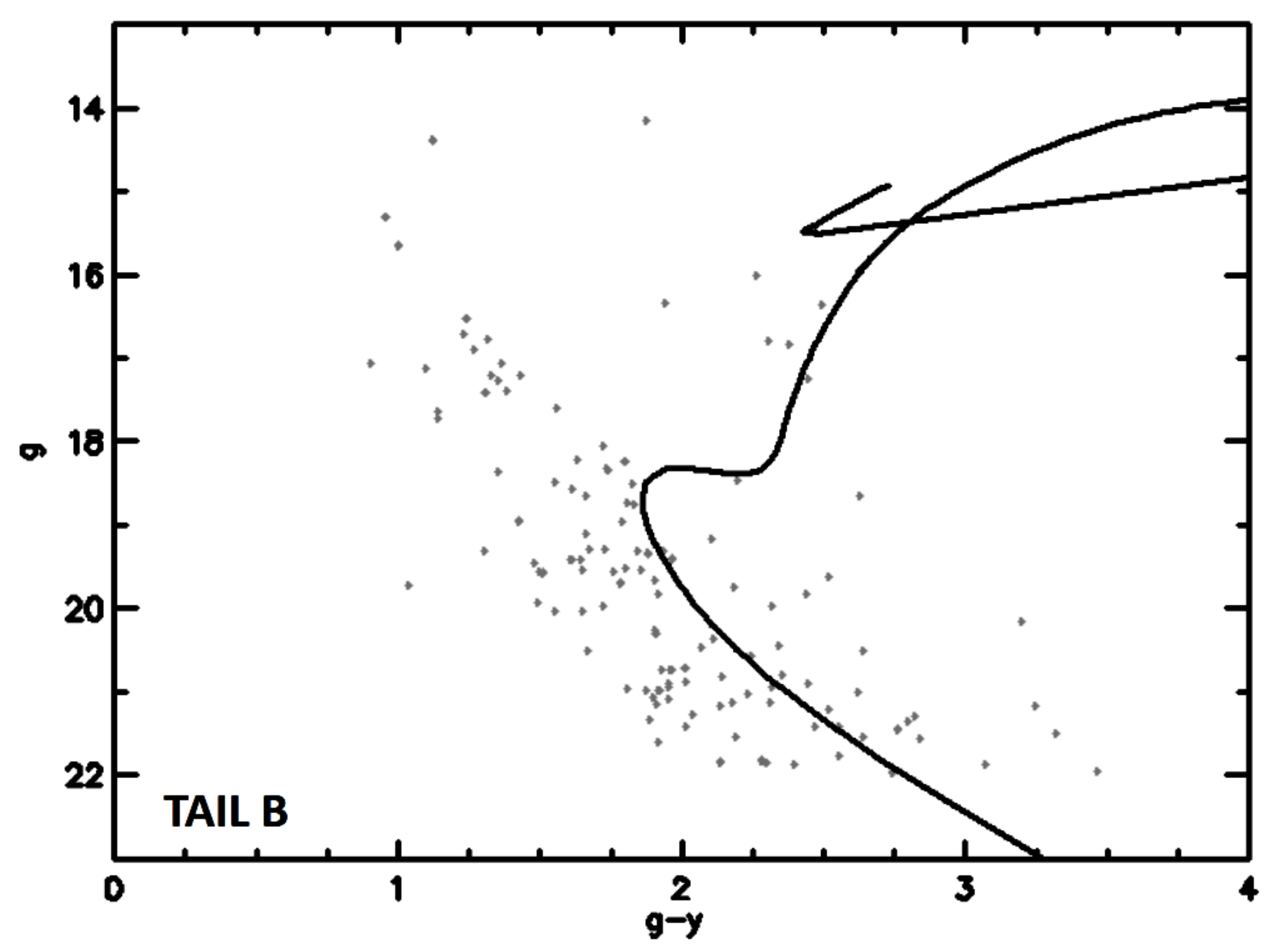}
	\caption{The observed CMDs for the core and the two tails. The blue straggler candidates listed by \citet{ahu07} have been encircled. The solid line marks the 10-Gyr isochrone, whereas the dashed line is a 100-Myr isochrone to illustrate the upper main sequence where blue stragglers are expected.}
	\label{fig:cmds}
\end{figure}

\begin{figure*}[htb]
	\centering
	\includegraphics[height=0.24\textheight,angle=0]{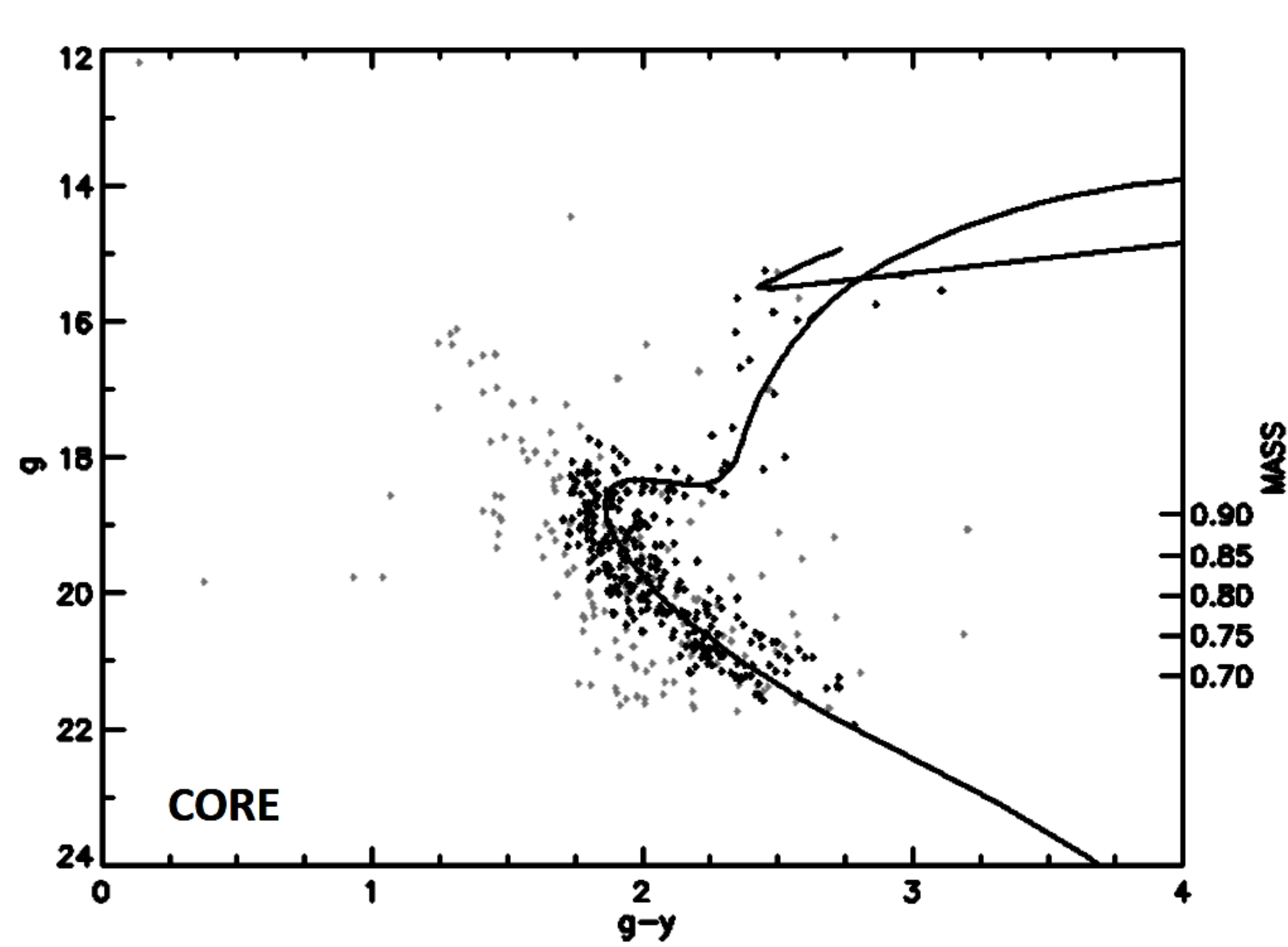}
	\includegraphics[height=0.24\textheight,angle=0]{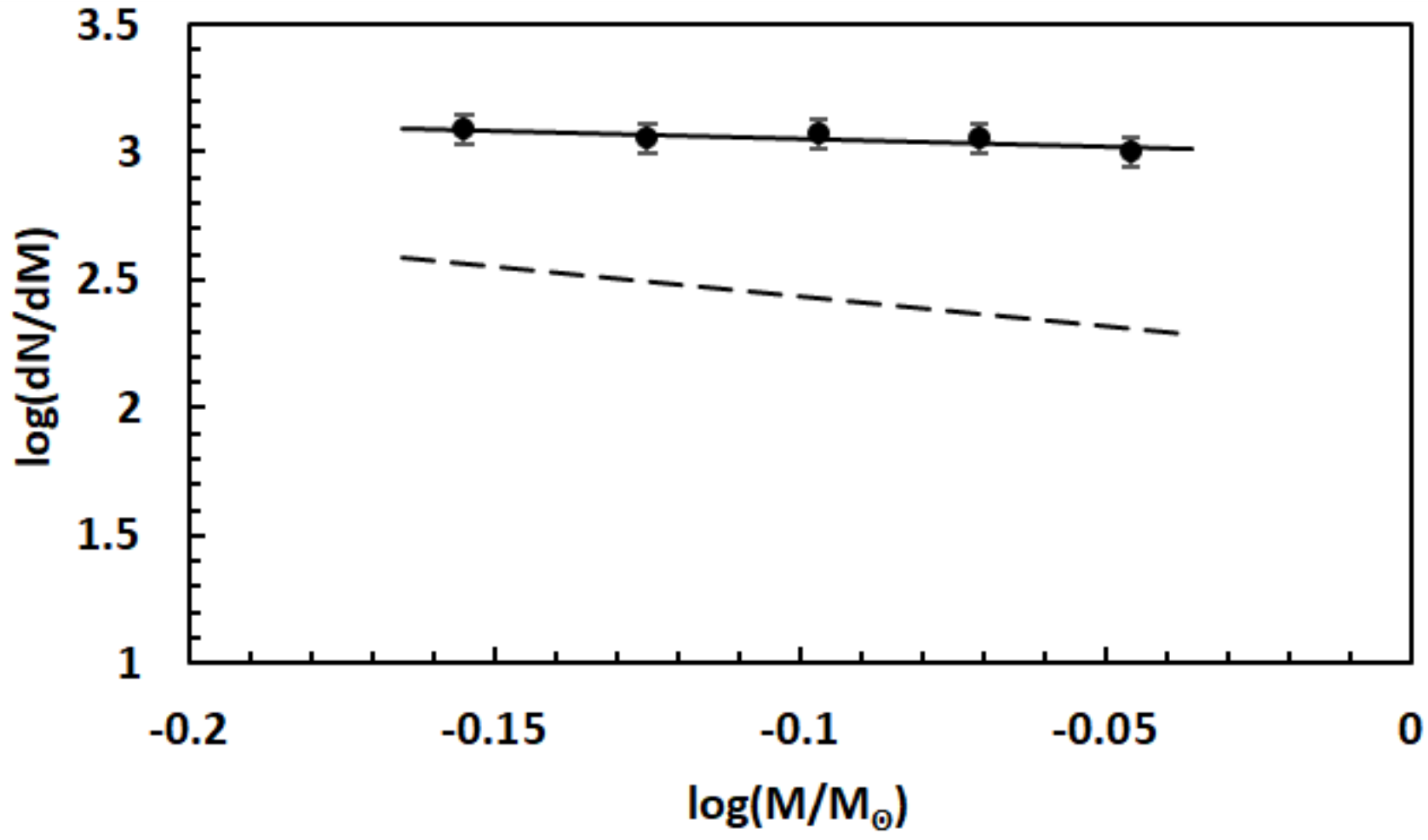}
	\includegraphics[height=0.24\textheight,angle=0]{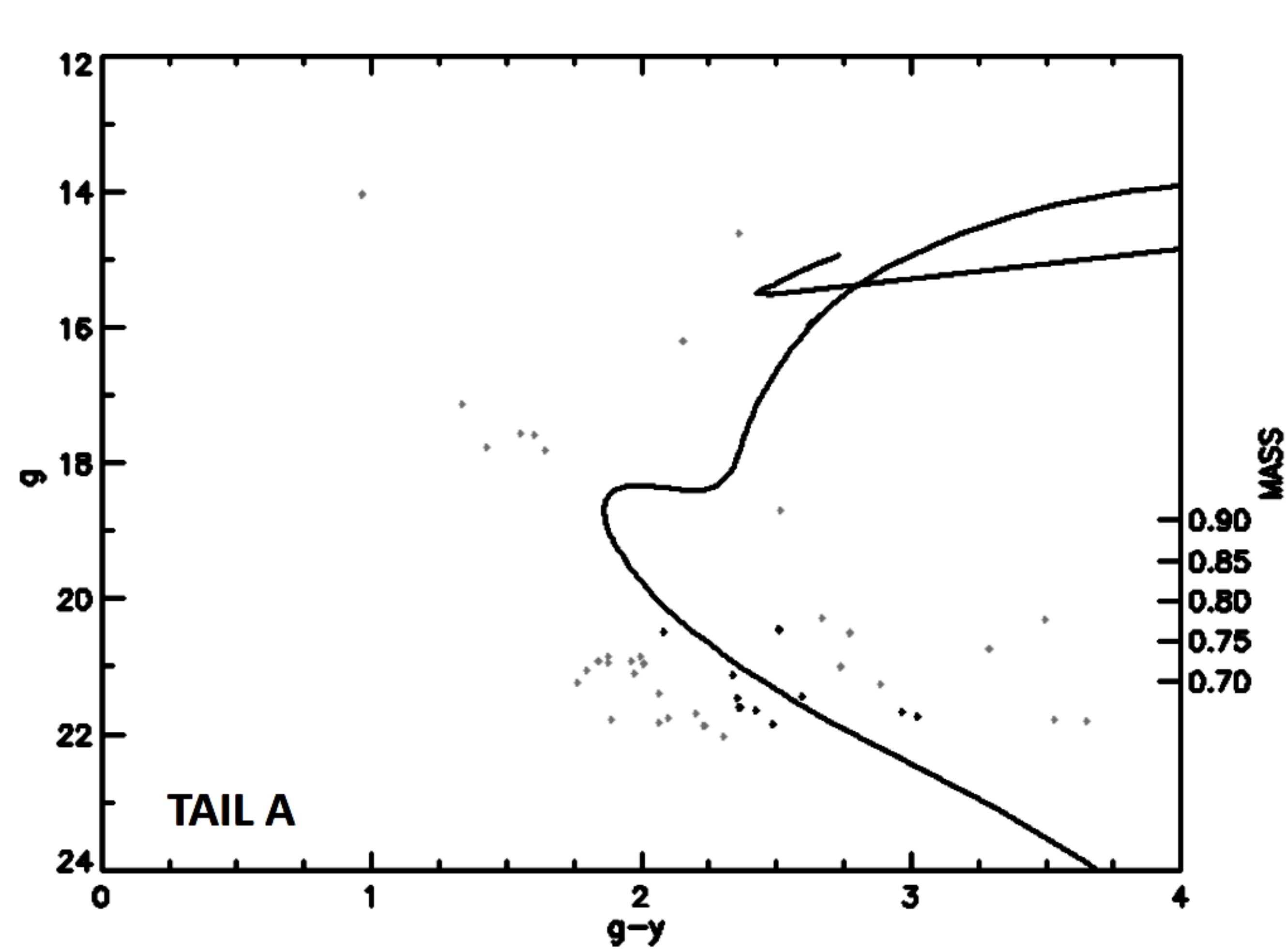}
	\includegraphics[height=0.24\textheight,angle=0]{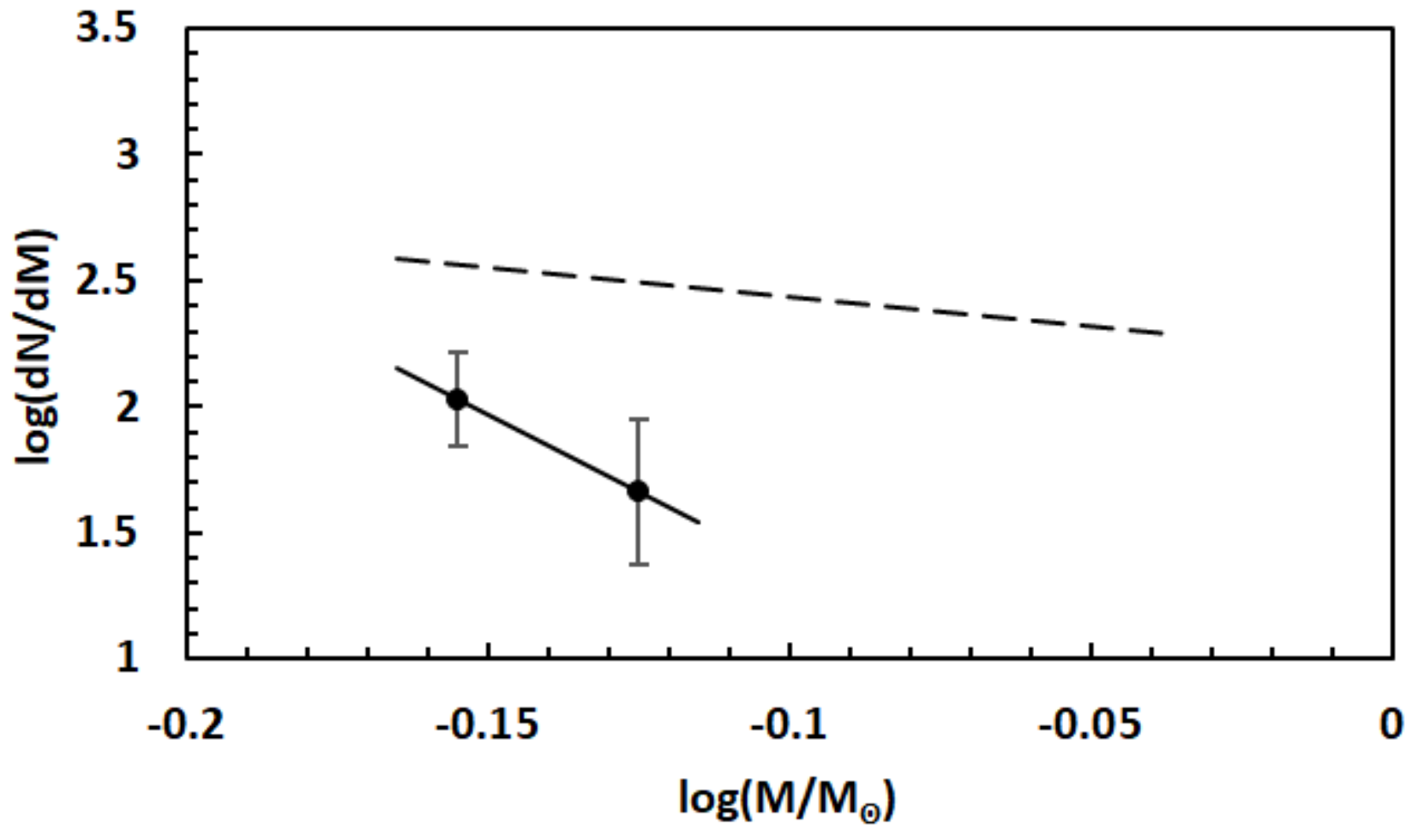}
	\includegraphics[height=0.24\textheight,angle=0]{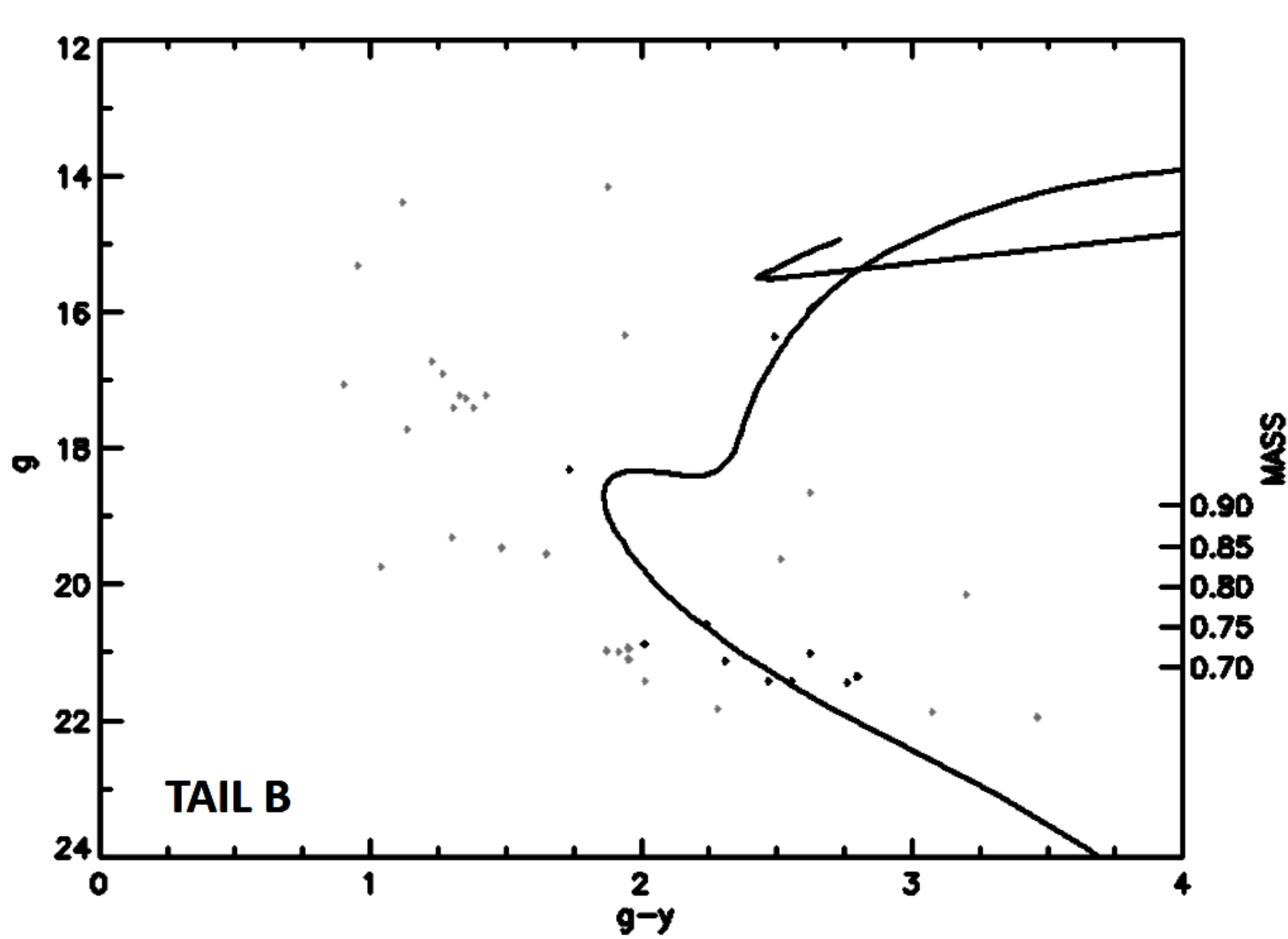}
	\includegraphics[height=0.24\textheight,angle=0]{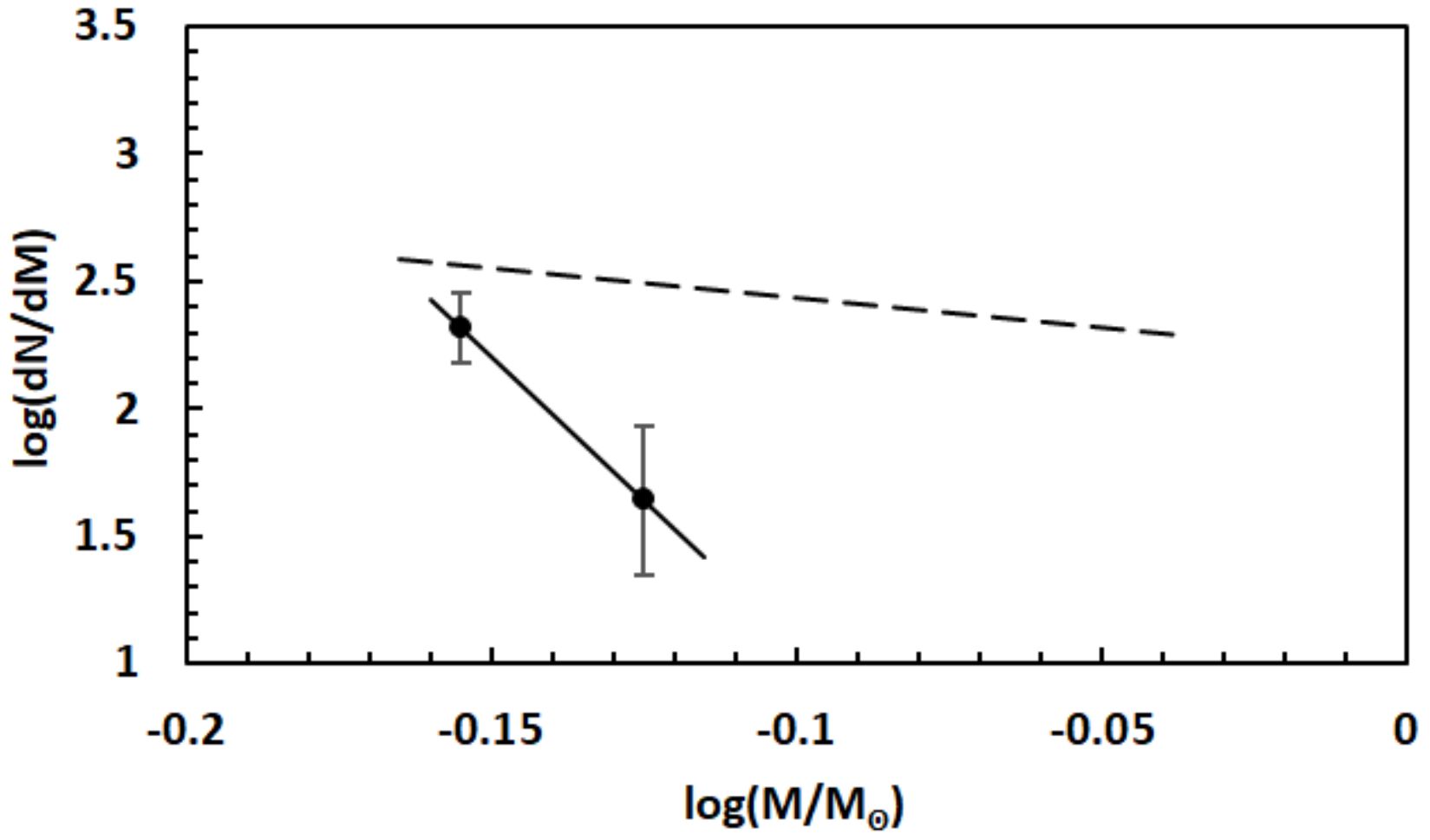}
	\caption{The cleaned CMDs and the isochrone members (in black) used to derive the corresponding mass functions, shown on the right, for the core and the two tails. The slope of the mass function for the core is $-0.39$. For the tails A and B, the slopes are $11.26$ and $21.45$ respectively, though the values are highly uncertain because of the small size of the sample. The Poisson uncertainties have been marked. The dashed line, shown for reference, corresponds to the mass function slope of $1.35$ derived by \citet{sal55} for field stars in the solar neighborhood.}
	\label{fig:massdist}
\end{figure*}

\subsection{Field Decontamination}\label{cmd}

To estimate field star contamination, we exercised a statistical ``cleaning" of the CMDs.  A reference field, having the same Galactic latitude as Berkeley\,17 and the same sky area as the cluster core, but located $1\degr$ away in longitude, was chosen.  For every star found in the field CMD (Figure~\ref{fig:fieldclean}), the star in the target CMD nearest to the field star would be eliminated if it was present within a box of size 0.56~mag in $g_{\rm P1} - y_{\rm P1}$ and 0.8~mag in $g_{\rm P1}$ centered on the field star. This ensured that each target star with a field counterpart would be eliminated, resulting in a cleaned CMD. The selection of the box size was somewhat subjective. It was chosen to account for the photometric uncertainties, and considered as the most suitable after trying either smaller ones, which proved ineffective in removing a sufficient number of field stars, or larger ones, which over-cleaned possible cluster members.  {Completeness was neglected during the field decontamination since any incompleteness is the same between the cluster and the field, as evident from Figure~\ref{fig:comp}.}

The effectiveness of our field decontamination procedure, and the random nature of the reference field, were affirmed by using another ``check field", also having the same Galactic latitude and sky area as the cluster core, to clean the reference field.  The result, depicted in Figure~\ref{fig:fieldclean}, was considered satisfactory as it is uncorrelated to the isochrone for Berkeley\,17. Therefore, the reference field was subsequently used to clean the core CMD. The CMDs for the tails were cleaned by the same manner, except with the reference field scaled down appropriately to the sky area of the tails.

Figure~\ref{fig:cmds} exhibits the observed CMDs for the core and the two tails, while the corresponding cleaned CMDs are shown in Figure~\ref{fig:massdist}. Stellar masses were inferred from the absolute $g_{\rm P1}$ magnitude, with a 10~Gyr PARSEC stellar evolution isochrone \citep{bre12,che14} for the PS1 filters, a distance of 2.7~kpc, and metallicity of 0.007 ( $[Fe/H] = -0.33$). The extinction values in the $g_{\rm P1}$ and $y_{\rm P1}$ bands were taken from those presented by \citet{sf11}, which are based on the reddening law from \citet{fit99}, assuming a nominal total-to-selective extinction $R_{\rm V}= 3.1$.  A reddening $E(B-V) = 0.6$ was found to have a marginally better isochrone fit to the main sequence than the alternative value of $E(B-V) = 0.7$ \citep{bra06,phe97}, but this had little effect on the mass estimation.

\subsection{Mass Distribution} \label{mdis}

We note that a cleaned CMD is a statistical representation of the cluster, so can be used to estimate the mass function, but the membership of individual stars is not available.
For our analysis, we only included stars with magnitudes below the turn off (no giants because the mass is uncertain) and within 0.14~mag in color of the isochrone (CMDs in Figure~\ref{fig:massdist}). This leads to a stellar mass range 0.675--0.925~M$_\sun$, with the lower limit corresponding to $g_{\rm P1} \sim~21.5$ mag. To determine the mass function, the number of members in each mass bin must be corrected for completeness as discussed in the Section~\ref{comp}. The mass functions of the core, and of both tails are presented in Figure~\ref{fig:massdist}. 

Even though our sample covers a relatively small mass range (0.25~M$_\sun$), the mass distribution in the core shows a paucity of lower-mass stars.  The result should not be due to observational bias, as photometric completeness has been taken into account, and is hence statistically significant even within the Poisson uncertainty, given a sufficient number of stars in each mass bin. The slope of the mass function $x$ for the cluster core (Figure~\ref{fig:massdist}), using the relation $\log(dN/dM) = -(1+x)\log(M) + \rm{constant}$, where $dN$ represents the number of stars in a mass bin $dM$ with central mass $M$, is $-0.39$.  This slope has an opposite sign in comparison to that of 1.35 derived by \citet{sal55} for field stars in the solar neighborhood. In contrast, the tails, albeit with a small sample in either case, consist exclusively of low-mass stars. Any observational bias, if present, would have been in favor of detection of high-mass stars.  {Tail A has a steep mass function slope of $11.26$ while that of Tail B is even steeper at $21.45$, though the values are highly uncertain because of the small size of the sample.} Thus, the stellar population in the tails is clearly different from that in the cluster core.

\subsection{Blue Straggler Candidates} \label{bst}

Blue stragglers have been studied in some open clusters for spectroscopic analysis \citep[NGC 2632]{and98}, in contact binaries \citep[NGC\,6791]{ruc96} or for variability \citep[Berkeley\,39]{kal93}. Some old open clusters, like globular clusters, are known to be rich in blue stragglers.  Berkeley\,17 was claimed to host some 31 blue straggler candidates \citep{ahu07} using the photometric data obtained by \citet{phe97}. We investigated these candidates with the same PS1 data used for the morphological study presented above. The equatorial coordinates of the candidates in \citet{ahu07} were made available from \citet{bra06} using WEBDA\footnote{https://www.univie.ac.at/webda}, and the PS1 counterparts were identified within $1\arcsec$, as presented in \citet{che17}. All of these candidates were found within our core region and their positions in the CMD are indicated in Figure~\ref{fig:cmds}. The aforementioned statistical cleaning resulted in the removal of 17 candidates, indicating about half of the candidates should likely be field contamination. Reliable proper motion and radial velocity measurements are needed to ascertain, or at least to exclude, their membership. 


\section{Discussion} \label{dis}

The metallicity of Berkeley\,17, ${\rm[Fe/H]}\approx-0.33$, is typical given its galactocentric distance $\sim11$~kpc \citep{car07} in the outer disk, where external perturbation is expected to be low. Still, the longevity of such an old open cluster remains puzzling. The size of the core, $\sim5$~pc, is within the ballpark figure for old clusters \citep{jan94}. While older open clusters typically have larger scale heights from the Galactic plane, Berkeley\,17 is close to the disk, with $z\sim170$~pc.  

The cause of the core-tail morphology of Berkeley\,17 is unclear.  With an antitail, it is likely of a tidal origin.  Combining the member samples in the core and in the tails, more massive stars outnumber the less massive ones, thus indicating a dynamically evolved state.  A flat mass function, in comparison to that of the solar neighborhood, found for Berkeley\,17 (see Figure~\ref{fig:massdist}) is not uncommon in old clusters \citep[for relevant references see][]{fri95}, as low-mass then-members have been stripped.  Our analysis did not include the post-main sequence population, but given $\sim250$ members in the 0.675--0.925~M$_\sun$ range, the original cluster should have been as massive as a super star cluster, commonly thought as the precursor of a globular cluster \citep{gg02}.  

One possible tidal source is the Perseus arm, known to be located at 1.95~kpc from the sun \citep{xu06}. \citet{kal94} attributed a distinct foreground field population toward Berkeley\,17 to belong to the Perseus arm. Indeed, such a foreground distribution also shows up in our observed CMDs, both for the cluster and for the field (see Figures~\ref{fig:fieldclean} and \ref{fig:cmds}). A sample of newly found star clusters toward the Galactic anticenter (Lin et al. in preparation) using PS1 data led to an estimated full-width-half-maximum of $\sim0.8\pm0.1$~kpc for the Perseus arm. Berkeley\,17 is immediately behind so potentially vulnerable to the tidal pull by the arm. One clue for the scenario is the unusual radial velocity for Berkeley\,17, $-84$~km~s$^{-1}$, indicating a large space $U$ velocity \citep{hay14}. 

The cleaned CMDs unequivocally show members above the main sequence turnoff. As presented above, roughly half of the blue straggler candidates \citep{ahu07} should be false positives. Some globular clusters \citep[e.g. NGC 3201;][]{bon10} have been seen to exhibit a prominent horizontal branch extending to the blue. The blue horizontal-branch stars have nearly constant brightness and so may serve as standard candles to constrain Galactic kinematics \citep{sir04a}. The blue horizontal branch may be related \citep{sw60} to, but cannot be entirely accounted for \citep{sw67,vdb67}, by metallicity. Other factors like stellar age and helium abundance have been proposed \citep{gra13,vil12}. In a CMD, some blue horizontal-branch stars overlap and thus may be confused with blue stragglers. Berkeley 17 is seen to have an extended horizontal branch \citep{che17}.  It is reasonable to speculate therefore that some of the blue straggler candidates may be blue horizontal-branch stars. Spectroscopic classification is required to clarify the nature of these blue stragglers versus blue horizontal-branch stars or field stars.

\acknowledgments

This work is financially supported by the grant MOST103-2112-M-008-024-MY3. The authors are grateful to the referee for the valuable comments. 
SB and IM express their gratitude to the Graduate Institute of Astronomy, NCU for the 
hospitality, support, and learning experience provided during their visit. 
This work made use of data products from the PS1 Surveys which have been made possible through contributions of the Institute for Astronomy, the University of Hawaii, the Pan-STARRS Project Office, the Max-Planck Society and its participating institutes, the Max-Planck Institute for Astronomy, Heidelberg and the Max-Planck Institute for Extraterrestrial Physics, Garching, The Johns Hopkins University, Durham University, the University of Edinburgh, Queen's University Belfast, the Harvard-Smithsonian Center for Astrophysics, the Las Cumbres Observatory Global Telescope Network Incorporated, the National Central University of Taiwan, and the National Aeronautics and Space Administration under grant no.\ NNX08AR22G issued through the Planetary Science Division of the NASA Science Mission Directorate. This research has made use of the WEBDA database, operated at the Department of Theoretical 
Physics and Astrophysics of the Masaryk University.

\end{document}